\documentclass[journal]{IEEEtran}
\usepackage{amsmath}
 \usepackage{graphicx} 
  \usepackage{epstopdf}
\begin{document}
\title{Outage Probability and Capacity Scaling Law of Multiple RIS-Aided Cooperative Networks}
\author{Liang Yang, Yin Yang, Daniel Benevides da Costa, and Imene Trigui
\thanks{L. Yang and Y. Yang are with the College of Information Science and Engineering, Hunan University, Changsha
410082, China, (e-mail:liangy@hnu.edu.cn, yy19971417@163.com).}
\thanks{D. B. da Costa is with the Department of Computer Engineering, Federal University of Cear\'{a}, Sobral, CE, Brazil (email: danielbcosta@ieee.org).}
\thanks{I. Trigui is with the University of Quebec, Montreal, QC, Canada (e-mail: trigui.imene17@gmail.com).}}
\maketitle
\begin{abstract}
In this letter, we consider a dual-hop cooperative network assisted by multiple reconfigurable intelligent surfaces (RISs). Assuming that the RIS with the highest instantaneous end-to-end signal-to-noise ratio (SNR) is selected to aid the communication, the outage probability (OP) and average sum-rate are investigated. Specifically, an exact analysis for the OP is developed. In addition, relying on the extreme value theory, closed-form expressions for the asymptotic OP and asymptotic sum-rate are derived, based on which the capacity scaling law is established. Our results are corroborated through simulations and insightful discussions are provided. In particular, our analysis shows that the number of RISs as well as the number of reflecting elements play a crucial role in the capacity scaling law of multiple RIS-aided cooperative networks. Also, comparisons with relay-aided systems are carried out to demonstrate that the proposed system setup outperforms relaying schemes both in terms of the OP and average sum-rate. 
\end{abstract}

\section{Introduction}
Reconfigurable intelligent surface (RIS) has been regarded as an emerging cost-effective technology for future wireless communication systems.
These artificial surfaces are composed of reconfigurable electromagnetic materials that can be controlled and programmed by integrated electronic devices, providing potential gains in terms of spectrum and energy efficiencies [1]. By comparing with traditional relaying technology, RISs do not need high hardware complexity and cost overhead during their operation. In addition, RISs are able to customize the wireless environment through the use of nearly-passive reflecting elements, enabling the system designers to fully control the electromagnetic response of the impinging signals into the environmental objects [2]. Therefore, due to their unique properties, RISs arise as one of the key technologies to realize the futuristic concept of smart radio environment.

Very recently, RISs have been extensively studied in the literature. Specifically, the authors in [3] formulated a beamforming optimization problem of RIS-aided wireless communication systems under discrete phase shift constraints. In [4], the authors applied RISs in downlink multi-user communications. The authors in [5] proposed a deep learning method for deploying RISs in an indoor environment. In [6], the authors quantitatively analyzed the coverage of RIS-assisted communication systems through an outage probability (OP) analysis. The authors in [7] conducted a performance analysis of the application of RISs in mixed free-space optical (FSO) and radio-frequency (RF) dual-hop communication systems. In [8], it was shown that the application of RISs can effectively improve the coverage and reliability of unmanned aerial vehicles (UAV) communication systems. In [9], the use of RISs for improving physical layer security was examined. On the other hand, along the years, the study of capacity scaling law of wireless communication systems has been of paramount importance. For example, in [10], it was studied the capacity
of multi-user multi-antenna relay networks with co-channel interference, while the authors in [11] analyzed the scaling rates for the OP and average sum-rate assuming a single RIS-assisted communication system.

Although the aforementioned works have provided interesting contributions, the research field on RIS is still at its infancy.
This paper aims to fill out an important gap which exists in the literature, which is the investigation of multiple RISs in dual-hop cooperative networks. To the best of the authors' knowledge, such kind of system setup has not been investigated in the literature yet.
Specifically, assuming a multiple RIS-aided system setup where the RIS with the highest instantaneous signal-to-noise ratio (SNR) is selected to assist the communication, an exact analysis for the OP is developed. In addition, relying on the extreme value theory, closed-form expressions for the asymptotic OP and asymptotic sum-rate are derived, based on which new capacity scaling laws are established. Our results are corroborated through simulations and insightful discussions are provided. In particular, our analysis shows that the number of RISs as well as the number of reflecting elements play a crucial role in capacity scaling law of multiple RIS-aided cooperative networks. Also, comparisons with relay-aided systems are carried out to show that the proposed system setup outperforms relaying schemes both in terms of the OP and average sum-rate.

The remainder of this paper is structured as follows. In Section II, the system and channel models are introduced. Section III carries out a detailed performance analysis in terms of the OP and average sum-rate, while Section IV presents illustrative numerical results which are followed by insightful discussions and corroborated through simulations. Finally, Section V concludes the paper.

\section{System and Channel Models}
As shown in Fig. 1, we consider a wireless communication system consisted of one source (S), one destination (D), and $K$ RISs, where each RIS is composed by $N$ reflecting elements. Similar to [1], it is assumed full channel state information (CSI). The communication process is briefly described next. Firstly, S sends the signal to the $k$-th RIS ($k=1,\ldots,K$)\footnote{As will be shown later, we consider that the RIS which provides the highest instantaneous end-to-end SNR is selected to assist the communication. However, at this moment, assume that the communication is carried out through the $k$-th RIS.} and then it passively reflects the signal to D\footnote{It is worth noting that RISs work similar to relaying systems, however their working principles are rather different and more detailed explanations can be found in [12].}. With this aim, the $k$-th RIS (denoted, for simplicity, by $\rm RIS_{\emph k}$) optimizes the phase reflection coefficient to maximize the received SNR at D, therefore improving the end-to-end quality of the communication system. The channels are assumed to undergo independent Rayleigh fading. Thus, the signal received at D can be expressed as
\begin{equation}
y_{Dk}=\sqrt{E_{s}}\left [ \sum_{i=1}^{N}h_{ki}e^{j\phi _{ki}} g_{ki}\right ]x+n,
\end{equation}
where $E_{s}$ denotes the average transmitted energy per symbol, $x$ is the transmitted signal, $ n\sim\mathcal{C}\mathcal{N}(0,N_{0})$ stands for
the additive white Gaussian noise (AWGN), $\phi_{ki}$ represents the adjustable phase produced by the $i$-th reflecting element of the $\rm RIS_{\emph k}$, $h_{ki}=d_{SR_{k}}^{-v/2}\alpha _{ki}e^{-j\theta_{ki}}$ and $g_{ki}=d_{R_{k}D}^{-v/2}\beta _{ki}e^{-j\varphi_{ki}}$ are the channel gains of the S-$\rm RIS_{\emph k}$ and $\rm RIS_{\emph k}$-D links, respectively,
where $d_{SR_{k}}$ and $d_{R_{k}D}$ denote the distances of the S-$\rm RIS_{\emph k}$ and $\rm RIS_{\emph k}$-D links, respectively, and $v$ denotes the path loss coefficient. In addition, $\alpha _{ki}$ and $\beta _{ki}$ represent the respective channels' amplitudes, which are independent distributed Rayleigh random variables (RVs) with mean $ \sqrt{\pi}/2$ and variance $(4-\pi)/4$, and $\theta_{ki}$ and $\varphi_{ki}$ refer to the phases of the respective fading channel gains. As considered in previous works, we assume that $\rm RIS_{\emph k}$ has perfect knowledge of $\theta_{ki}$ and $\varphi_{ki}$. From (1), the instantaneous end-to-end SNR at D can be expressed as
\begin{equation}
\gamma _{k}=\frac{E_{s}\left | \sum_{i=1}^{N}\alpha _{ki} \beta _{ki}e^{j(\phi _{ki}-\theta_{ki}-\varphi_{ki})}\right|^{2}}{N_{0}d_{SR_{k}}^{v}d_{R_{k}D}^{v}}.
\end{equation}
\begin{figure}
\centering
\includegraphics[height=4cm,width=8cm]{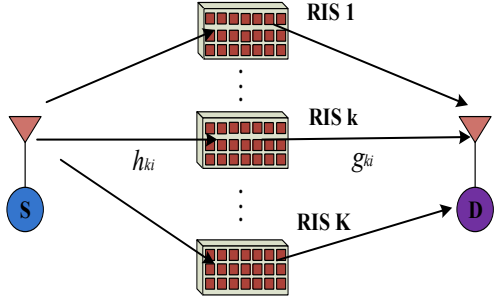}
\caption{System model.}
\end{figure}

From [1], in order to maximize $\gamma _{k}$, $\rm RIS_{\emph k}$ can be smartly configured to fully eliminate the phase shifts by setting $\phi _{ki}=\theta_{ki}+\varphi_{ki}$. Therefore, the maximized $\gamma _{k}$ can be written as
\begin{equation}
\gamma _{k}=\frac{E_{s}\left (\sum_{i=1}^{N} \alpha _{ki} \beta _{ki} \right )^{2}}{N_{0}d_{SR_{k}}^{v}d_{R_{k}D}^{v}}={\bar{\gamma}_{k}A^{2}},
\end{equation}
where $A=\sum_{i=1}^{N}\alpha _{ki} \beta _{ki}$ and $ \bar{\gamma}_{k}=\frac{E_{s}}{N_{0}d_{SR_{k}}^{v}d_{R_{k}D}^{v}}$ stands for the average SNR.

\section{Performance Analysis}
In this section, we analyze the OP and average sum-rate of the considered system setup. Along the analytical derivations, it will be considered that one out of $K$ RIS is selected to aid the communication. Specifically, the choice of the suitable RIS is performed to  maximize the received signal at the destination. Therefore,
the selection principle of the RIS can be expressed as
\begin{equation}
k^{\ast}={\rm arg}\max \limits_{k\in\{1,2,\ldots,K\}}\frac{\gamma_{k}}{\tilde{\gamma}_{k}},
\end{equation}
where $\tilde{\gamma}_{k}$ is the average channel SNR measured by $\rm RIS_{\emph k}$ in the past window of length $s_{t}$. Note that $\tilde{\gamma}_{k}$ is introduced in the denominator to maintain the long-term fairness. Next, similar to [13] ,we focus on the investigation of small-scale channel fading. Thus,
the selection principle in (4) simplifies to
\begin{equation}
k^{\ast}={\rm arg}\max \limits_{k\in\{1,2,\ldots,K\}}\gamma_{k}.
\end{equation}

In what follows, the analysis is developed assuming a clustered configuration for the deployment of the RISs, which implies that the end-to-end links undergo independent identically distributed (i.i.d.) Rayleigh fading\footnote{Non-clustered configuration analysis arises as an interesting study for future works. Therefore, the results obtained in this paper can serve as a benchmark for these new investigations.}.

\subsection{Outage Probability Analysis}
\subsubsection{ Exact Analysis}
Based on the clustered deployment of the RISs, it follows that $\gamma_{1}=\gamma_{2}=\ldots=\gamma_{k}=\gamma$, which yields
$A=\sum_{i=1}^{N}\alpha _{i} \beta _{i}$. Let $B_{i}=\alpha _{i} \beta _{i}$, then the probability density function (PDF) of $B_{i}$ can be readily obtained as $f_{B_{i}}(\gamma)=4\gamma K_{0}(2\gamma)$, where $K_{0}(\cdot)$ is the modified Bessel function of the second kind
with zero order [14]. According to [15], one can attest that the PDF of $B_{i}$ is a special case of the $K_{G}$ distribution. In [15], the authors stated that the PDF of the sum of multiple $K_{G}$ RVs can be well-approximated by the PDF of $\sqrt{W}$ with $W=\sum_{i=1}^{N}B_{i}^{2}$, in which the PDF of $W$ is approximated by a squared $K_{G}$ distribution. Therefore, the PDF of $A^{2}$ can be represented by a squared $K_{G}$  distribution.
Thus, the PDF of $\gamma$ can be written as [16]
\begin{equation}
f_{\gamma}(\gamma)=\frac{2\Xi^{l+m}}{\Gamma(l)\Gamma(m)\bar{\gamma}^{\frac{l+m}{2}}}\gamma^{(\frac{l+m}{2}-1)}K_{l-m}(2\Xi\sqrt{\gamma/\bar{\gamma}}),
\end{equation}
where $l$ and $m$ are the shaping parameters, $\Gamma(\cdot)$
denotes the gamma function, $\Omega\overset{\triangle }{=}E[A^{2}]$ is
the mean power, and $\Xi=\sqrt{lm/\Omega}$.

Relying on the idea presented in [17], we make use of the mixed gamma (MG) distribution to rewrite the PDF given in (6) as
\begin{align}
f_{\gamma}(\gamma)=\sum_{i=1}^{M}w_{i}\gamma^{\rho_{i}-1}\bar{\gamma}^{-\rho_{i}}e^{-\varepsilon_{i}\gamma/\bar{\gamma}},
\end{align}
where $M$ denotes the number of terms of the sum, $w_{i}=\frac{\chi_{i}}{\sum_{j=1}^{M}\chi_{j}\Gamma(\rho_{j})\varepsilon_{j}^{-\rho_{j}}}$,
$\rho_{i}=m$, $\varepsilon_{i}=\Xi^{2}/t_{i}$, and $\chi_{i}=\frac{\Xi^{2m}y_{i}t_{i}^{l-m-1}}{\Gamma(m)\Gamma(l)}$,
with $y_{i}$ and $t_{i}$ representing, respectively, the weight factor and the abscissas of the Gaussian-Laguerre integration [18]. From probability theory, the cumulative distribution function (CDF) of $\gamma$ can be derived as
\begin{equation}
F_{\gamma}(\gamma)=\sum_{i=1}^{M}w_{i}\varepsilon_{i}^{-\rho_{i}}\Upsilon\left(\rho_{i}, \frac{\varepsilon_{i}\gamma}{\bar{\gamma}}\right),
\end{equation}
where $\Upsilon(\cdot,\cdot)$ represents the lower incomplete gamma function [14].

According to the order statistics theory, the CDF of $\gamma_{k^{\ast}}$, which is given in (5), can be formulated as
\begin{equation}
F_{\gamma_{k^{\ast}}}(\gamma)=(F_{\gamma}(\gamma))^{K},
\end{equation}
which yields the following PDF
\begin{equation}
f_{\gamma_{k^{\ast}}}(\gamma)=Kf_{\gamma}(\gamma)[F_{\gamma}(\gamma)]^{K-1}.
\end{equation}

From [19], the OP can be defined as the probability that the effective received SNR $\gamma_{k^{\ast}}$ is less than a given threshold $\gamma_{\rm th}$, which is mathematically written as $P_{out}=\Pr(\gamma_{k^{\ast}}<\gamma_{\rm th})$. Thus, by replacing (8) into (9), the system OP can be expressed as
\begin{equation}
P_{out}=\left[\sum_{i=1}^{M}w_{i}\varepsilon_{i}^{-\rho_{i}}\Upsilon\left(\rho_{i},\frac{\varepsilon_{i}\gamma_{\rm th}}{\bar{\gamma}}\right)\right]^{K}.
\end{equation}
\subsubsection{Asymptotic Analysis}
According to [14], $\Upsilon(a,b)$ can be rewritten in the form $\Upsilon(a,b)=e^{-b}\sum_{n=0}^{\infty}(b^{a+n}/a(a+1)\ldots(a+n))$.
Based on this identity, $\Upsilon(\rho_{i},\varepsilon_{i}\gamma/\bar{\gamma})$ can be asymptotically expressed as
\begin{align}
\Upsilon\left(\rho_{i},\frac{\varepsilon_{i}\gamma}{\bar{\gamma}}\right)\simeq e^{-\frac{\varepsilon_{i}\gamma}{\bar{\gamma}}}\left(\rho_{i}^{-1}
\left(\frac{\varepsilon_{i}\gamma}{\bar{\gamma}}\right)^{\rho_{i}}+o(\gamma^{\rho_{i}+1})\right).
\end{align}
At SNR regime, one can ignore the high order term $o(\gamma^{\rho_{i}+1})$. Thus, an asymptotic outage expression for (11) can be expressed as
\begin{equation}
P_{out}\simeq\frac{1}{[\bar{\gamma}^{\rho_{i} }]^{K}}\left[\sum_{i=1}^{M}w_{i}e^{-\frac{\varepsilon_{i}\gamma_{\rm th}}{\bar{\gamma}}}\rho_{i}^{-1}\gamma_{\rm th}^{\rho_{i}}\right]^{K}.
\end{equation}
The above expression indicates that the achievable diversity order of the proposed system setup is $K\rho_{i}$, which can also be written as   $KN$ since $\rho_{i}$ equals to $m$, and this latter is determined by $N$.

\subsection{Asymptotic Sum-Rate Analysis}
To analyze the asymptotic sum-rate, we depart from the idea presented in [20, Lemma 2]. More specifically, let $\{z_{1}$, \ldots,  $z_{K}\}$ i.i.d. RVs with a common CDF $F_{Z}(\cdot)$ and PDF $f_{Z}(\cdot)$, satisfying the property that $F_{Z}(\cdot)$ is less than one for all finite $z$ and is twice differentiable for all $z$, which implies that
\begin{equation}
\lim_{z \to \infty  }\frac{1-F_{Z}(z)}{f_{Z}(z)}=C> 0.
\end{equation}
for some constant $C$. Then, the expression $\max \limits_{1\leq k\leq K}z_{k}-h_{K}$ converges in distribution to a limiting RV with CFD given by $\rm exp(-e^{-\emph x/C})$. It is worth noting that the CDF of $h_{K}$ is given by $1-\frac{1}{K}$.

The above result allows to say that the maximum of $K$ i.i.d. RVs grows like $h_{K}$. Therefore, in the sequel we derive the asymptotic sum-rate  assuming a high number of RISs, i.e., as $K\longrightarrow\infty$. Due to the intricacy in carrying out such kind of analysis departing from the MG distribution presented in the previous section, here, for sake of tractability, the CDF of $\gamma$ will be written by using the non-central chi-square distribution [6], i.e.,
\begin{equation}
F_{\gamma}(\gamma)=1-Q_{\frac{1}{2}}\left(\frac{\sqrt{\lambda}}{\sigma},\frac{\sqrt{\gamma}}{\sqrt{\bar{\gamma}}\sigma}\right),
\end{equation}
where $Q_{\nu}(c,d)$ denotes the Marcum $Q$-function, $\lambda=(\frac{N\pi}{4})^2$, and $\sigma^2=N(1-\frac{\pi^2}{16})$.
At high SNR regime, based on [22], the Marcum $Q$-function $Q_\tau(x,y)$ can be asymptotically expressed as
\begin{align}
Q_\tau(x,y)\simeq(1-2\vartheta)^{-\tau}\exp(-\vartheta \emph y^2)\exp\left(\frac{\tau\vartheta x^2}{1-2\vartheta} \right),
\end{align}
where $\vartheta$ stands for the Chernoff parameter $(0<\vartheta<\frac{1}{2})$. By replacing (16) into (15), the CDF of $\gamma $ can  be asymptotically expressed as
\begin{align}
F_{\gamma}(\gamma)&\simeq 1-(1-2\vartheta )^{-\frac{1}{2}}\exp\left(-\frac{16\vartheta\gamma }{\bar{\gamma }\emph N(16-\pi ^{2})}\right)
\nonumber\\&\times\exp\left(\frac{\vartheta \emph N\pi ^{2}}{2(1-2\vartheta )(16-\pi ^{2})}\right),
\end{align}
and its corresponding PDF is
\begin{align}
f_{\gamma}(\gamma)&\simeq (1-2\vartheta )^{-\frac{1}{2}}\frac{16\vartheta }{\bar{\gamma }N(16-\pi ^{2})}\exp\left(-\frac{16\vartheta\gamma  }{\bar{\gamma }\emph N(16-\pi ^{2})}\right)\nonumber\\
&\times\exp\left(\frac{\vartheta \emph N\pi ^{2}}{2(1-2\vartheta )(16-\pi ^{2})}\right).
\end{align}

Then, we can show that
\begin{equation}
\lim_{\gamma \to \infty  }\frac{1-F_{\gamma}(\gamma)}{f_{\gamma}(\gamma)}=\frac{\bar{\gamma }N(16-\pi ^{2})}{16\vartheta }=C_{1}> 0.
\end{equation}

Also, by solving $F(h_{K})=1-\frac{1}{K}$, it follows that
\begin{equation}
\begin{aligned}
h_{K}=&\left [\ln\emph K-\frac{1}{2}\rm  ln(1-2\vartheta) +\frac{\vartheta \emph N\pi ^{2}}{2(1-2\vartheta )(16-\pi ^{2})}\right ]\\
&\times\frac{\bar{\gamma }N(16-\pi ^{2})}{16\vartheta }.
\end{aligned}
\end{equation}

Therefore, for a large number of RISs, the maximum SNR $\gamma$ grows as in (20), which is the function of $K$ and $N$ for fixed $\bar{\gamma}$. Accordingly, the asymptotic sum-rate can be approximated
by
\begin{equation}
\begin{aligned}
C^{K}\simeq  &\log_{2}(1+\emph h_{\emph K})\\
\simeq  &\log_{2}\left(\ln\emph K-\frac{1}{2}\rm ln(1-2\vartheta)+\frac{\vartheta \emph N\pi ^{2}}{2(1-2\vartheta )(16-\pi ^{2})}\right)\\
+&\log_{2}\left(\frac{\bar{\gamma }N(16-\pi ^{2})}{16\vartheta }\right)\\
\simeq  &\log_{2}(\ln\emph K)+\log_{2}\left(\frac{16-\pi^{2}}{16\vartheta}\right)+\log_{2}\bar{\gamma}+\log_{2}\emph N.
\end{aligned}
\end{equation}

The above expression indicates that $K$ and $N$ play a significant role in increasing the sum-rate.

\section{Numerical Results and Discussions}
In this section, illustrative numerical examples are presented to verify the impact of the key system parameters on the overall performance.
Our analysis is corroborated through Monte Carlo simulations, in which $10^5$ simulation points are generated. Moreover, comparisons with relay-aided systems are carried out to demonstrate that the proposed system setup outperforms relaying schemes both in terms of the OP and average sum-rate.

In Fig. 2, the OP of RIS-aided systems and relaying systems is plotted assuming $\gamma_{\rm th}=20\, \rm dB$, $K=1,2,3$, and $N=3$. From this figure, one can see that RIS-aided systems have better performance than relaying ones. Moreover, the system performance improves as $K$ increases, in which the slope of the curves changes according to $K$, corroborating the presented asymptotic analysis. In Fig. 3, the OP of RIS-aided systems for different values is plotted for different values of $K$ and $N$. It can be clearly seen that increasing the value of $N$ can significantly improve the system performance. In both figures, one can observe that the analytical results match perfectly with the simulation results. Also, at high SNRs, the asymptotic results are close to the exact values.
\begin{figure}
\centering
\includegraphics[width=\linewidth]{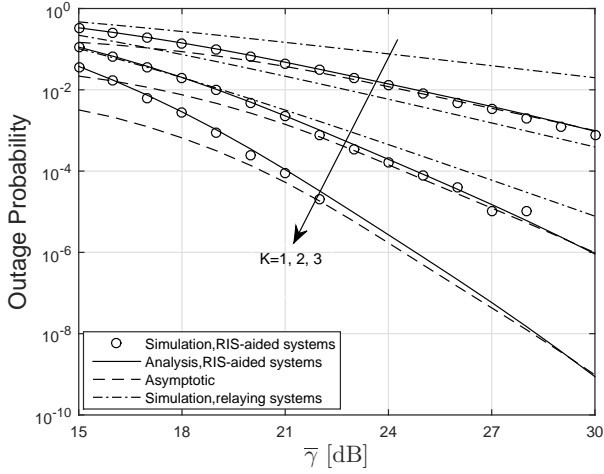}
\caption{Outage probability versus $\bar{\gamma}$ for different number of RISs/relays and assuming $N=3$.}
\end{figure}
\begin{figure}
\centering
\includegraphics[width=\linewidth]{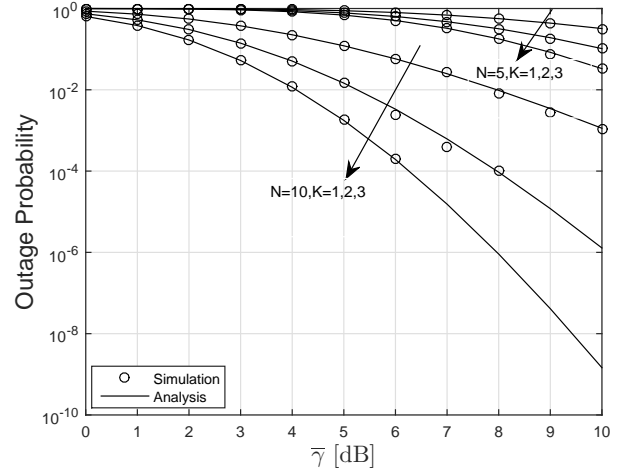}
\caption{Outage probability versus $\bar{\gamma}$ for different values of $K$ and $N$.}
\end{figure}

Fig. 4 depicts the OP for different combinations of $N$ and $K$, but keeping the product $KN$ with the same value. It can be clearly seen that the four curves have the same slopes, which verifies the correctness of the derived diversity order, i.e., $KN$. Furthermore, the slight change of $N$ and $K$ can lead to an improved system performance, however such performance is mostly dependent on the value of $N$.
\begin{figure}
\centering
\includegraphics[width=\linewidth]{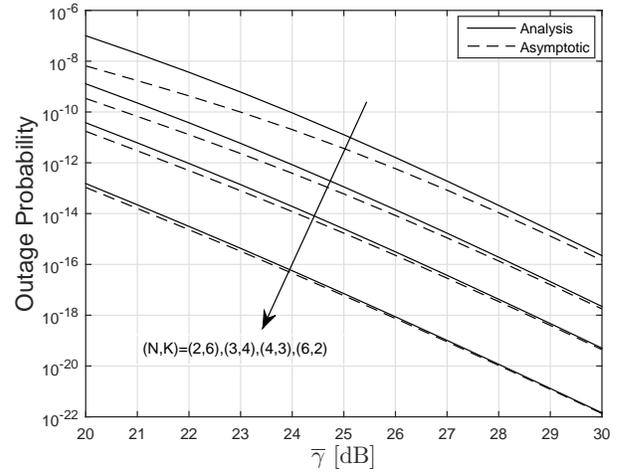}
\caption{Outage probability for different combinations pairs of ($N$, $K$).}
\end{figure}

In Fig. 5, the asymptotic sum-rate is plotted for different values of $N$ and assuming $K = 5$. From this figure, one can attest that $N$ has a great impact on the system performance. Our results are also compared with the relay schemes (assuming the same number as considered for RIS-aided case, i.e., $K=5$) to show the performance gain of the proposed system setup. In Fig. 6, the average sum-rate versus $K$ is depicted by setting $N=10,15$ and $\gamma_{\rm th} =10\, \rm dB$. It can be clearly seen that the asymptotic curves well reflect the scaling law of the considered RIS-aided system, although they do not coincide with simulation results. However, the asymptotic value becomes tighter to the simulation values  when $N$ increases. Finally, it can be observed that the proposed system setup is significantly better than the relay schemes in terms of average sum-rate.
\begin{figure}
\centering
\includegraphics[width=\linewidth]{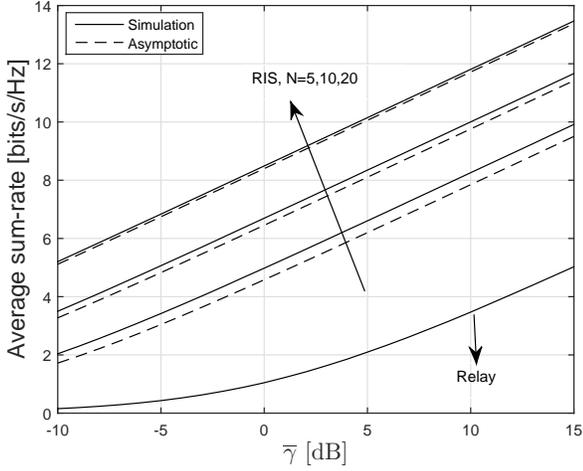}
\caption{Average capacity of RIS-aided systems for different values of $N$ and assuming $K=5$ RISs/relays.}
\end{figure}
\begin{figure}
\centering
\includegraphics[width=\linewidth]{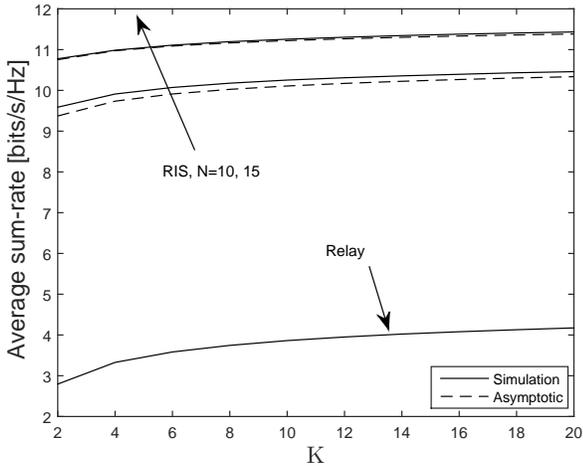}
\caption{Average capacity versus $K$ for different values of $N$.}
\end{figure}
\section{Conclusion}
\vspace{-0.1cm}
In this work, we investigated dual-hop cooperative network assisted by multiple RISs. Assuming that one of out $K$ RISs (which one having $N$ elements) is selected to add the communication process, an exact analysis for the OP was presented and closed-form expressions for the asymptotic OP and asymptotic sum-rate were derived, based on which the capacity scaling law was determined. Our results showed that the achievable diversity order of the considered system equals to $KN$. In addition, we have compared the proposed scheme with a DF relay scenario and it is shown that the former one outperforms considerably the latter one both in terms of OP and average sum-rate. Our results are not only novel but can be used as a benchmark for future studies. Potential new works include non-clustered configuration for the RISs' deployment as well as the proposal of other techniques of RIS selection.

 \appendices
\ifCLASSOPTIONcaptionsoff
  \newpage
\fi


\begin{thebibliography}{99}
\bibitem{IEEEhowto:kopka}
E. Basar, M. Di Renzo, J. De Rosny, M. Debbah, M. S. Alouini and R. Zhang, ``Wireless communications through reconfigurable intelligent surfaces," \emph{IEEE Access}, vol. 7, pp. 116753-116773, Aug. 2019.
\bibitem{IEEEhowto:kopka}
C. Liaskos, S. Nie, A. Tsioliaridou, A. Pitsillides, S. Ioannidis, and I.
Akyildiz, ``A new wireless communication paradigm through software-
controlled metasurfaces," \emph{IEEE Commun. Mag.},  vol. 56, no. 9, pp. 162-
169, Sep. 2018.
\bibitem{IEEEhowto:kopka}
Q. Wu and R. Zhan, ``Beamforming optimization for wireless network aided by intelligent reflecting surface with discrete phase shifts," \emph{IEEE
Trans. Wireless Commun.}, vol. 68, no. 3, pp. 1838-1851, Mar. 2020.
\bibitem{IEEEhowto:kopka}
C. Huang, A. Zappone, G. C. Alexandropoulos, M. Debbah and C. Yuen, ``Reconfigurable intelligent surfaces for energy efficiency in wireless
communication," \emph{IEEE Trans. Wireless Commun.}, vol. 18, no. 8, pp. 4157-4170, Aug. 2019.
\bibitem{IEEEhowto:kopka}
C. Huang, G. C. Alexandropoulos, C. Yuen, and M. Debbah, ``Indoor
signal focusing with deep learning designed reconfigurable intelligent
surfaces," in\emph{ Proc. IEEE IWSPAWC}, Cannes, France, 2019,
pp. 1-5.
 \bibitem{IEEEhowto:kopka}
L. Yang, Y. Yang, M. O. Hasna, and M. Alouini, ``Coverage, prob-
ability of SNR gain, and DOR analysis of RIS-aided communication
systems," \emph{IEEE Wireless Commun. Lett.}, Early Access, DOI: 10.1109/L-
WC.2020.2987798.
 \bibitem{IEEEhowto:kopka}
 L. Yang, W. Guo, and I. S. Ansari, ``Mixed dual-hop FSO-RF communi-
cation systems through reconfigurable intelligent surface," \emph{IEEE Commun.
Lett.},  vol. 24, no. 7, pp. 1558-1562, Jul. 2020,
 \bibitem{IEEEhowto:kopka}
L. Yang, F. Meng, J. Zhang,  M. O. Hasna, and M. Di Renzo, ``On the performance of RIS-assisted dual-hop UAV communication systems," \emph{IEEE Trans.  Veh. Technol.}, Early Access, DOI:10.1109/TVT.2020.3004598.
  \bibitem{IEEEhowto:kopka}
L. Yang, J. Yang, W. Xie, M. O. Hasna, T. Tsiftsis, M. Di Renzo, ``Secrecy performance analysis of RIS-aided wireless communication systems," \emph{IEEE Trans.  Veh. Technol.}, Early Access, DOI: 10.1109/TVT.2020.3007521.

 \bibitem{IEEEhowto:kopka}
I. Trigui, S. Affes and A. St¨¦phenne, ``Capacity scaling laws in interference-limited multiple-antenna AF relay networks with user scheduling," \emph{IEEE Trans. Commun.}, vol. 64, no. 8, pp. 3284-3295, Aug. 2016.
 \bibitem{IEEEhowto:kopka}
I. Trigui, W. Ajib and W. Zhu, ``A comprehensive study of reconfigurable intelligent surfaces in generalized fading," [Online]. Available: https://arxiv.org/abs/2004.02922.

 \bibitem{IEEEhowto:kopka}
  K. Ntontin, M. Di Renzo, J. Song, F. Lazarakis, J. D. Rosny, D.
T. Phan Huy, O. Simeone, R. Zhang, M. Debbah, G. Lerosey, M.
Fink, S. Tretyakov, S. Shamai, ``Reconfigurable intelligent surfaces
vs. relaying: Differences, similarities, and performance comparison," [Online]. Available: https://arxiv.org/abs/1908.08747.


\bibitem{IEEEhowto:kopka}
 C. Chen and L. Wang, ``A unified capacity analysis for wireless systems with joint multiuser scheduling and antenna diversity in Nakagami fading channels," \emph{IEEE Trans. Commun.}, vol. 54, no. 3, pp. 469-478, Mar. 2006.
\bibitem{IEEEhowto:kopka}
I. S. Gradshteyn, and I. M. Ryzhik, \emph{Table of integrals, series, and
products}, 7th ed. San Diego, CA, USA: Academic, 2007.
\bibitem{IEEEhowto:kopka}
 K. P. Peppas, ``Accurate closed-form approximations to generalised-$K$
sum distributions and applications in the performance analysis of equal-
gain combining receivers," \emph{IET Commun.}, vol.5, no.7, pp. 982-989, May
2011.
\bibitem{IEEEhowto:kopka}
L. Yang, F. Meng, Q. Wu, D. B. da Costa and M. Alouini, ``Accurate closed-form approximations to channel distributions of RIS-aided wireless systems," \emph{IEEE Wireless Commun. Lett.}, Early Access, DOI:10.1109/LWC.2020.3010512.


\bibitem{IEEEhowto:kopka}
S. Atapattu, C. Tellambura, and H. Jiang, ``A mixture gamma distribution
to model the SNR of wireless channels," \emph{IEEE Trans. Wireless Commun.}, vol. 10, no. 12, pp. 4193-4203, Dec. 2011.
\bibitem{IEEEhowto:kopka}
 M. Abramowitz and I. A. Stegun, \emph{Handbook of mathematical functions:
With formulas, graphs, mathematical tables.} New York, USA:
Dover, 1965.


 \bibitem{IEEEhowto:kopka}
 M. K. Simon and M.S. Alouini, \emph{Digital communication over fading
channels: A unified approach to performance analysis}, 1st Edition.
John Wiley, 2000.
\bibitem{IEEEhowto:kopka}
 P. Viswanath, D. N. C. Tse, and R. Laroia, ``Opportunistic beamforming
using dumb antennas," \emph{IEEE Trans. Inform. Theo.},  vol. 48, pp.
1277-1294, Jun. 2002.
\bibitem{IEEEhowto:kopka}
 V. M. Kapinas, S. K. Mihos, and G. K. Karagiannidis, ``On the
monotonicity of the meneralized marcum and nuttall Q-functions," \emph{IEEE Trans. Inform. Theo.},  vol. 55, no. 8, pp. 3701-3710, Aug. 2009.
\bibitem{IEEEhowto:kopka}
  M. K. Simon and M. S. Alouini, ``Exponential-type bounds on the
generalized marcum Q-function with application to error probability
analysis over fading channels," \emph{IEEE Trans. Commun.},  vol. 48, no. 3,
pp. 359-366, Mar. 2000.


\end{thebibliography}
\end{document}